\documentstyle [12pt] {article}
\begin{document}


\pagestyle{empty}
\rightline{\vbox{
\halign{&#\hfil\cr
&hep-ph/9405348\cr
&UCD-94-2\cr
&May 1994\cr}}}
\bigskip
\bigskip
\bigskip
{\bf \centerline{Perturbative QCD Fragmentation Functions for Production of}
	\centerline{$P$--wave Mesons with Charm and Beauty}}
\bigskip
\normalsize

\centerline{Tzu Chiang Yuan}
\centerline{\sl Davis Institute for High Energy Physics}
\centerline{\sl Department of Physics, University of California,
    Davis, CA  95616}
\bigskip

\begin{abstract}

We calculate the leading order QCD
fragmentation functions for the production of $P$-wave charmed beauty
mesons.
Long-distance effects are factored into two nonperturbative parameters:
the derivative of the radial wavefunction at the origin and a second
parameter related to the probability for a $(\bar b c)$ heavy quark pair
that is produced in a color-octet $S$-wave state to form a color-singlet
$P$-wave bound state.
The four $2P$ states and those $3P$ states which lie
below the $BD$ flavor threshold
eventually all decay into the $1S$ ground state $B_c$ through
hadronic cascades or by emitting photons.
The total fragmentation probabilities for production of the
$1S$ ground state $B_c$
from the cascades of the $2P$ and $3P$ states
are about $1.7 \times 10^{-4}$ and $2.3 \times 10^{-4}$
respectively.
Thus the direct production of the $P$-wave states via
fragmentation may account for a significant fraction of the
inclusive production rate of the $B_c$ at large
transverse momentum in high energy colliders.
Our analytic results for  the $P$-wave fragmentation functions disagree
with those obtained earlier in the literature.
Our results also cover the $P$-wave heavy
quarkonium case in the equal mass limit.

\end{abstract}

\vfill\eject\pagestyle{plain}\setcounter{page}{1}

\section{Introduction}

Much progress has been made in the past several years
to improve our theoretical understanding
of the physics of a heavy hadron containing a single heavy quark
(charm or bottom quark). This
is mainly due to the discover of the powerful
heavy quark spin-flavor symmetry \cite{wiseguys} in
Quantum Chromodynamics (QCD) and the
development of the Heavy Quark Effective Theory  \cite{hqet}.
Heavy hadrons containing both the heavy bottom and charm quarks are also
of interest. Of particular importance is the $(\bar b c)$ meson family.
Since the top quark is very heavy and will decay rapidly
into $W^+b$ before hadronization, the $(\bar b c)$ system
might be the only physical system containing two heavy flavors
of different masses. Techniques of Heavy Quark Effective Theory can not be
applied here because the charm quark is not sufficient light.
However, since this family is
intermediate between the $c \bar c$ charmonium
and the $b \bar b$ bottomonium systems, many properties
(mass spectrum, transition and decay rates, lifetimes, {\it etc}.)
of this physical system can be studied by using
potential models and QCD sum rules  \cite{eichtenquigg,baganetal}.

Recently, it has been pointed out \cite{bcfrags}
that the dominant production mechanism for these yet unobserved
$(\bar b c)$ bound states with large transverse momentum
at a high energy process
is fragmentation, in which an energetic $\bar b$ antiquark is first
produced at large transverse momentum by a short-distance
hard process and subsequently
fragments into various $(\bar b c)$ bound states.  Furthermore,
the process independent fragmentation functions describing these
phenomena are shown to be calculable at the heavy quark mass scale
using perturbative QCD. In particular,
the fragmentation functions for the $\bar b$ antiquark to split
into the two $S$-wave states $B_c(n^1S_0)$ and
$B_c^*(n^3S_1)$ were calculated \cite{bcfrags,changchen}
to leading order in both $\alpha_s$ and $v$,
where $v$ is the typical relative velocity of the charm quark
inside the meson.
Including the $1S$ and $2S$ states that lie below the $BD$ flavor threshold
$(M_{\rm threshold} = M_D + M_B \approx 7.1 \; {\rm GeV})$,
a lower bound on the inclusive branching fraction for the production
of the $B_c$ ground state from $Z^0$ decay has been estimated to be about
$2.3 \times 10^{-4}$ \cite{bcfrags}, which might be too small
for this particle to be observed at LEP with the present luminosity.
Production of the $B_c$ meson at the hadron colliders via the direct
$\bar b$ antiquark and induced gluon
fragmentation has also been studied in Ref.\cite{bctevatron}.
About 20,000 $B_c$ with transverse momentum $p_T > 10 $ GeV
are expected to be produced at the Tevatron
with an integrated luminosity of $25 \, {\rm pb}^{-1}$.
A clean signature of $B_c$ would be the
observation of three charged leptons coming
from the same secondary vertex due to
the decay $B_c^+ \to J/\psi + {\bar l}' \nu_{l'}$ followed by
$J/\psi \to l \bar l$.  The combined branching fraction for
these decays is expected to
be about $0.2 \%$ \cite{branching}.
One expects about 40 of these distinctive events at the Tevatron.
Thus, unless LEP can increase its luminosity
by an order of magnitude or so in the near future,
the best place to look for the charmed beauty meson will be at the Tevatron.

The mass spectrum of the $(\bar b c)$ mesons has been obtained by
Eichten and Quigg \cite{eichtenquigg} using potential model calculations.
Since, unlike the quarkonium system, the $\bar b$ and $c$ quarks do not
annihilate into gluons, the excited states of the
$(\bar b c)$ mesons can not decay directly into light hadrons.
States that lie above the $BD$ threshold will strongly decay into
a pair of $B$ and $D$ mesons. States that lie below the $BD$ threshold
will decay eventually to the ground state
$B_c$ either by hadronic cascades or by emitting  photons.
In order to get more reliable theoretical prediction for the
inclusive production rate of the $B_c$ meson at large transverse
momentum,  one must include all higher orbitally excited
states of the $(\bar b c)$ mesons that lie below the $BD$ threshold.
Below the $BD$ flavor threshold, there are two sets
of $S$-wave states (the $1S$ and $2S$ states),
as many as two complete sets of $P$-wave states
(the $2P$ and perhaps some or all $3P$ states),
and one set of $D$-wave states (the $3D$ states).

In this paper, we
report the results of the calculation of the fragmentation functions
for $\bar b$ antiquark splittings into the
four $P$-wave $(\bar b c)$ mesons --
the $^1P_1$ and $^3P_J$ $(J=0,1,2)$ states.
In Section 2, we write down new factorization formulas for the
$P$-wave fragmentation functions that are valid to all orders in strong
coupling constant $\alpha_s$ and to leading order in $v$, where
$v$ is the typical relative velocity of the heavy quarks
inside the $(\bar b c)$ meson.
Long-distance effects are factored into two nonperturbative
parameters: the derivative of the radial wavefunction at the origin, and
a second parameter related to the probability for the $(\bar b c)$ pair that
is produced in a color-octet $S$-wave state to form a color-singlet $P$-wave
bound state.
In Section 3, we calculate the perturbative coefficients of the fragmentation
functions to leading order in $\alpha_s$.
Infrared logarithmically-divergent terms that appeared at
the next-to-leading order
in $\alpha_s$ are isolated and absorbed into the second
nonperturbative parameter.  Mixing effects between the $^1P_1$ and $^3P_1$
states in the fragmentation functions are also calculated.
In Section 4, we take the heavy quark limit
of $m_b/m_c \rightarrow \infty$, and show that our fragmentation functions
exhibit the heavy quark spin symmetry while earlier results obtained by
Chen \cite{pwave_chen} do not satisfy this important property.
In Section 5, we discuss a crossing
relation between the fragmentation function $D_{i \to H}(z)$ for parton $i$
splittings into hadron $H$ and the distribution function $f_{i/H}(z)$
of finding parton $i$ inside hadron $H$.
In Section 6, we consider the equal mass limit corresponding to
the case of $P$-wave heavy quarkonium. Numerical results and
conclusions are made in Section 7.

\bigskip

\section{New factorization formulas for $P$-wave fragmentation functions}

Recently Bodwin, Braaten, and Lepage \cite{bbl_long} have reformulated the
perturbative calculations of the annihilation rates
of heavy quarkonium using the framework of non-relativistic QCD
(NRQCD) \cite{bbl_long,NRQCD}.
This work points out that the usual factorization assumption that all
long-distance effects can be absorbed into the radial wavefunctions at the
origin and their derivatives is not correct when higher order QCD corrections
and/or relativistic corrections are taken into account.
Infrared divergences that arise in these higher order calculations spoil
factorization.
To obtain sensible perturbative results that are free from infrared
divergences, one must use new factorization formulas that involve a
double expansion in the strong coupling constant $\alpha_s$
and in the typical velocity $v$ of the
heavy quark inside the quarkonium \cite{bbl_long}. The easiest way to
organize the expansion in $v$ in these theoretical calculations
is to use NRQCD \cite{bbl_long,NRQCD}.

As in the quarkonium case \cite{gfragp},
the heavy quarks inside the $(\bar b c)$ meson are moving nonrelativistically
and separated at a typical distance of order $1/(mv)$, where $v \ll 1$ is the
typical relative velocity of the heavy quarks
and $m=m_bm_c/(m_b+m_c)$ is the reduced mass.
The calculation of the perturbative fragmentation function is based on
separating short-distance perturbative effects
involving the scale of order $1/m$ from long-distance nonperturbative
effects involving scales of order $1/(mv)$ or larger.
To leading order in $v$, two distinct mechanisms that contribute to the
fragmentation function have been identified and are referred as the
{\it color-singlet mechanism} and {\it color-octet mechanism}. The
{\it color-singlet mechanism} is the production of a $(\bar b c)$ pair in a
color-singlet $^1P_1$ or $^3P_J$ state with separation of order $1/m$ in the
$(\bar b c)$ meson rest frame. The subsequent formation of the
$P$-wave $(\bar b c)$ bound state is a long-distance nonperturbative process
with a probability of order $v^2 (m v/m)^3 \sim v^5$, where the first factor
$v^2$ arises from the derivative of the wavefunction of the $P$-wave state and
the second factor $(mv/m)^3$ is a volume factor \cite{bbl_long,bbl_short}.
The {\it color-octet mechanism} is the production of the
$(\bar b c)$ pair in a color-octet
$^1S_0$ or $^3S_1$ state with a separation of
order $1/m$. The subsequent formation of the
$P$-wave $(\bar b c)$ bound state can proceed either through the
dominant $|\bar b c \rangle$ component or through the small
$|\bar b c g \rangle$ component of the wavefunction.
In the first case, the $(\bar b c)$ pair must radiate a soft gluon to make a
transition to the color-singlet $P$-wave bound state. In the second case, a
soft gluon must combine with the $(\bar b c)$ pair to form the color-singlet
$P$-wave bound state. In both cases, the probability is of order $v^5$, with
a factor of $v^3$ coming from the volume factor and an addition factor
of $v^2$ coming either from the probability of radiating a soft gluon or from
the small probability of the $|\bar b c g \rangle$ component of the
wavefunction \cite{bbl_long,bbl_short}.
Since the {\it color-singlet mechanism} and
{\it color-octet mechanism} contribute to the
fragmentation function at the same order in $v$, they must both be
included for a consistent calculation.

Separation of the short-distance physics from the long-distance effects
discussed in the previous paragraph requires the introduction of an arbitrary
factorization scale $\Lambda$ in the range
$m v \ll \Lambda \ll m$. The fragmentation functions for the
$P$-wave $(\bar b c)$ mesons satisfy factorization formulas that
involve this arbitrary scale.
To leading order in $v^2$ and to all order in $\alpha_s$,
the factorization formulas for the fragmentation functions of
$\bar b \to \bar b c(n^1P_1)$ and
$\bar b \to \bar b c(n^3P_J)$ at a scale $\mu_0$ near the heavy quark mass
scale consist of two terms:
\begin{eqnarray}
\label{d1p1total}
D_{\bar b \to \bar b c(n^1P_1)}(z,\mu_0) & = & {H_{1(\bar bc)}(n) \over m}
D^{(1)}_{\bar b \to \bar b c(^1P_1)}(z,\Lambda) +
3 {H'_{8(\bar b c)}(\Lambda) \over m}
D^{(8)}_{\bar b \to  \bar b c(^1S_0)}(z) \; , \\
\label{d3pjtotal}
D_{\bar b \to \bar b c(n^3P_J)}(z,\mu_0) & = & {H_{1(\bar b c)}(n) \over m}
D^{(1)}_{\bar b \to \bar b c(^3P_J)}(z,\Lambda)
+ (2J+1) {H'_{8(\bar b c)}(\Lambda) \over m}
D^{(8)}_{\bar b \to \bar b c(^3S_1)}(z) \; ,
\end{eqnarray}
where $H_1$ and $H_8'(\Lambda)$ are nonperturbative long-distance factors
associated with the {\it color-singlet} and {\it color-octet mechanism}
respectively, $m$ is the reduced mass,
and $n=2,3,\cdots$, labels the principal quantum number
of the $P$-wave states.
The short-distance factors $D^{(1)}(z,\Lambda)$ and $D^{(8)}(z)$ can be
calculated perturbatively as power series in $\alpha_s$.
They are proportional to the fragmentation functions for a $\bar b$ antiquark
to split into a $(\bar b c)$ pair with vanishing
relative momentum and definite color-spin-orbital quantum numbers:
color-singlet $^1P_1$ or $^3P_J$ state for $D^{(1)}$ and
color-octet $^1S_0$ or $^3S_1$ state for $D^{(8)}$. Note that in these
factorization formulas, the only dependence on $\Lambda$ is in $D^{(1)}$ and
$H_8'$.
As in the quarkonium case, the nonperturbative parameters $H_1$ and $H'_8$
can be rigorously defined as matrix elements of 4-quark operators
in nonrelativistic QCD divided
by appropriate powers of the heavy quark masses \cite{bbl_long}.
Their dependence on $\Lambda$ is governed by a renormalization group equation
whose coefficient can be calculated using nonrelativistic QCD \cite{bbl_long}.
We adopt the same definitions for these matrix elements as in the
charmonium case discussed in Ref.\cite{bbl_long}, except that explicit factors
of $m_c$ are replaced by $2m$.
To order $\alpha_s$, $H_{1(\bar b c)}(n)$ is scale invariant
and is related to  the derivative of the
non-relativistic radial wavefunction at the origin for the $P$-wave
$(\bar b c)$ bound states:
\begin{equation} {
H_{1(\bar b c)}(n) \; \approx \; {9 \over 2\pi}
{|R'_{nP}(0)|^2 \over (2m)^4} \;
\left( 1 + {\cal O}(v^2) \right) \; ;
} \label{hone} \end{equation}
while $H'_{8(\bar b c)}$ satisfies \cite{bbl_long,bbly}
\begin{equation} {
\Lambda {d \ \over d \Lambda} H'_{8(\bar b c)}(\Lambda)
\; = \; {16 \over 27\pi}
	\alpha_s(\Lambda)H_{1(\bar b c)} \; ,
} \label{rgeq} \end{equation}
with the solution
\begin{equation} {
H'_{8(\bar b c)}(\Lambda) \;  = \; H'_{8(\bar b c)} (\Lambda_0)
\; + \; {16 \over 27 \beta_0}
	\log \left( {\alpha_s(\Lambda_0) \over \alpha_s(\Lambda)} \right)
\; H_{1(\bar b c)} \; ,
} \label{rgsol} \end{equation}
where $\beta_0 = 9/2$ is the first coefficient in the beta
function for QCD with 3 flavors of light quarks.
If the factorization scale $\Lambda$ is chosen to be much less than
the reduced mass $m$,
the above equation can be used to sum up large logarithms of $m/\Lambda$.
To avoid large logarithms in the perturbative coefficients
$D^{(1)}$ in Eqns.(\ref{d1p1total}) and (\ref{d3pjtotal}), one
can choose $\Lambda$ on the order of $m$.
Since $|R'_{nP}(0)|^2$ is of order $v^5$ \cite{bbl_long},
from Eqns.(\ref{hone}) and (\ref{rgsol}),
we see that the {\it color-singlet} and
{\it color-octet} contributions in the fragmentation functions
are both of order $v^5$. Following Ref.\cite{bcfrags}, the initial
scale $\mu_0$ for the $P$-wave fragmentation functions in
Eqns.(\ref{d1p1total}) and (\ref{d3pjtotal})
can be chosen
to be $(m_b + 2m_c)$ -- the minimal virtuality
of the fragmenting $\bar b$ antiquark.
Fragmentation functions at a higher scale
can be obtained by solving the Altarelli-Parisi evolution equation with
Eqns.(\ref{d1p1total}) and (\ref{d3pjtotal}) as the boundary conditions.
These boundary conditions will be determined in the next Section.

\bigskip

\section{Fragmen\-tation func\-tions for $\bar b$ $\rightarrow$
$P$-wave $(\bar b c)$ me\-sons}

We now embark on the calculation of the coefficient
$D^{(1)}(z,\Lambda)$ in the color-singlet contribution to the
fragmentation function.
Let us first briefly review the method that was introduced
in Refs.\cite{bcfrags,gfrags} of calculating the
heavy quark fragmentation functions using perturbative QCD.
We refer to Ref.\cite{bcfrags} for more details.
Let ${\cal M}$ denote the
amplitude for a high energy source (symbolically denoted by $\bar \Gamma$)
to create a $\bar b^*(q) \rightarrow H(p) + \bar c(p')$
with total 4-momentum $q = p+p'$ as illustrated in Fig.1. Here
$H$ denotes a $(\bar b c)$ bound state.  The leading order
diagram of Fig.1 involves creating a $(c \bar c)$ pair from the vacuum.
Let ${\cal M}_0$ denote the
amplitude for the same source to create an on-shell $\bar b$ antiquark
with the same 3-momentum $\vec q$. Then the fragmentation function for
$\bar b \rightarrow H$ is given by \cite{bcfrags}
\begin{equation}
D_{\bar b \rightarrow H}(z) =
{1 \over 16\pi^2} \int ds \; \theta \Biggl (s - {(m_b + m_c)^2 \over z}
- {m_c^2 \over 1-z} \Biggr) \lim_{q_0/m_{b,c} \rightarrow \infty}
{|{\cal M}|^2 \over |{\cal M}_0|^2} \, ,
\label{dz}
\end{equation}
where $s = q^2$ is the virtuality of the fragmenting $\bar b$ antiquark.
In a frame where the virtual $\bar b$ antiquark has 4-momentum
$q = (q_0,0,0,q_3)$, the longitudinal momentum fraction of $H$
is $z=(p_0+p_3)/(q_0+q_3)$ and its transverse momentum is
$\vec p_{\bot} = (p_1,p_2)$. The relation between the transverse momentum
and the virtuality is given by
\begin{equation}
\vec p_{\bot}^{\; 2} = z(1-z) \Biggl( s - {M^2 \over z} - {r^2M^2 \over 1-z}
			\Biggr) \; .
\label{ptsz}
\end{equation}
The tree-level matrix element squared $|{\cal M}_0|^2$ is simply given by
$ 3 {\rm Tr} (\Gamma \bar \Gamma q\llap/)$
as $q_0/m_b \rightarrow \infty$.
The matrix element ${\cal M}$ for producing a bound state $H(p)$
consisting of collinear $\bar b$ and $c$
quarks with relative momentum $k$ (Fig.1)
can be expressed as a Dirac trace involving the spinor factor
$v(\bar r p + k) \bar u (r p - k)$. To project out the $(\bar b c)$ bound
state in the non-relativistic approximation,
this spinor factor is replaced by  the following projection operator
\begin{equation}
v(\bar r p + k) \bar u (r p - k) \to
\sqrt{m_b+m_c}
\left( {\bar r \not\! p \, + \not\!k - m_b \over 2m_b} \right)
{\cal O}
\left( {r \not\! p \, - \not\!k + m_c \over 2m_c} \right) \; ,
\end{equation}
with ${\cal O} = \gamma_5 $ or --$\epsilon\llap/^*(p,S_z)$ for
the spin singlet or spin
triplet state respectively, where $\epsilon(p,S_z)$ is the
polarization vector associated with the spin triplet state.
Besides this spinor factor, the gluon propagator in Fig.1 also depends on the
relative momentum $k$. In the axial gauge associated with the
4-vector $n^\mu$, the gluon propagator is given by
\begin{equation}
D^{\mu\nu}_F(l) =
{1 \over l^2} \Bigg[
-g^{\mu\nu}+{n^\mu l^\nu + n^\nu l^\mu \over n \cdot l}
+{n^2 l^\mu l ^\nu \over (n \cdot l)^2} \Bigg]  \; ,
\end{equation}
where $l =  q - \bar r p - k$. The axial gauge vector can be
chosen to be $n^\mu = (1,- \vec p / |\vec p|)$ so that Fig.1 is the dominant
diagram and hence factorization is manifest.
By using the equations of motion for the $\bar b$ and
$c$ quarks, one can show that terms that are
proportional to $l^\nu$ in the gluon
propagator do not contribute to the matrix element ${\cal M}$.
For the $P$-wave states, one needs to expand the above projection
operator and the gluon propagator appearing in the
matrix element ${\cal M}$ to first order in the relative momentum $k$.
After some manipulations by
using the standard covariant methods \cite{kuhnetal},
the matrix element for the $n^1P_1$ state can be written
in the following form:
\begin{equation}
{\cal M} (n^1P_1) = \delta_{ij} {1 \over 3r^2\bar r}
{g^2 R^{\prime}_{nP} (0) \over \sqrt{4\pi M}}  {1 \over (s -\bar r^2 M^2)^3}
\epsilon^*_{\alpha}(p,L_z) \bar \Gamma V^{\alpha} \gamma_5 v (p') \, ,
\label{onepone}
\end{equation}
where $i$ and $j$ are the fundamental color indices,
$g$ is the strong coupling
constant, and $\epsilon_{\alpha}(p,L_z)$ is the polarization vector of the
$^1P_1$ state. The vertex factor $V^{\alpha}$ is
given by
\begin{eqnarray}
V^\alpha & = & 8\bar r q^\alpha (\not \! q + \bar r M)(\not \! p - 2M)
\nonumber\\
& - & (s - \bar r^2M^2)\Bigg[ 2(1-2r)(\not \! q+\bar rM)\gamma^\alpha
  - {4 \bar r \over n \cdot (q-\bar r p)} q^\alpha (\not \! p+M){\not \! n}
\Bigg]
\nonumber\\
& - & {(s - \bar r^2M^2)^2 \over n \cdot (q-\bar r p)}
\Bigg[ {1 \over M} (\not \! p + (1-2r)M) \gamma^\alpha
		- 2r \bar r {1 \over n \cdot (q - \bar r p)}
		n^\alpha (\not \! p + M) \Bigg]{\not \! n} \, .
\label{tensor1}
\end{eqnarray}
In Eqn.(\ref{tensor1}), we have defined
$M=m_b+m_c$, $r=m_c/M$, and $\bar r = 1-r = m_b/M$.
Similarly, the matrix elements ${\cal M}(n^3P_J)$ for
the three spin triplet states can be written in the following compact form:
\begin{equation}
{\cal M} (n^3P_J) = \delta_{ij} {1 \over 3r^2\bar r}
{g^2 R^{\prime}_{nP} (0) \over \sqrt{4\pi M}} {1 \over (s -\bar r^2 M^2)^3}
\Psi_{\alpha\beta}(J,J_z) {\bar \Gamma} V^{\alpha\beta}  v(p') \; ,
\label{matrix3pj}
\end{equation}
with
\begin{eqnarray}
\Psi_{\alpha\beta}(J,J_z) & = &
\sum_{L_z,S_z} \langle 1,L_z;1,S_z|J,J_z \rangle
\epsilon^*_\alpha(p,L_z)\epsilon^*_\beta(p,S_z) \\
& = & \left \{ \begin{array}{ll}
{1 \over \sqrt 3}(g_{\alpha\beta} - {p_\alpha p_\beta \over M^2})
& \mbox{for $J=0,$} \\
{i \over \sqrt 2}{1 \over M} \epsilon_{\alpha\beta\xi\eta}
p^\xi \epsilon^{\eta *}(p,J_z)
& \mbox{for $J=1,$} \\
\epsilon_{\alpha\beta}(p,J_z)
& \mbox{for $J=2,$} \end{array} \right.
\label{polar}
\end{eqnarray}
where
$\epsilon_\alpha (p,L_z)$ and $\epsilon_\beta (p,S_z)$ are the polarization
vectors in the orbital and spin space of the $^3P_J$ states respectively;
and
$\epsilon_\alpha (p,J_z)$ and $\epsilon_{\alpha\beta}(p,J_z)$
are the helicity wave functions of the total spin
$J=1$ and $J=2$ states respectively.
Our conventions are $g_{\mu\nu}={\rm diag}(1,-1,-1,-1)$ and
$\epsilon^{0123}=+1$.
The vertex tensor $V^{\alpha\beta}$ in Eqn.(\ref{matrix3pj}) is given by
\begin{eqnarray}
V^{\alpha\beta} & = & 8\bar r M q^\alpha (\not \! q + \bar r M)\gamma^\beta
\nonumber\\
& + & (s - \bar r^2M^2) \Bigg[ {1 \over M}(\not \! q+\bar rM)
\left(
2 g^{\alpha\beta} ((1-2r)\not \! p-2M) +
[\gamma^\alpha,\gamma^\beta] \not \! p
\right)
\nonumber\\
& + & {4 \bar r \over n \cdot (q-\bar r p)} q^\alpha
(\not \! p+M) \gamma^\beta {\not \! n}  \Bigg]
\nonumber\\
& + & {(s - \bar r^2M^2)^2 \over n \cdot (q-\bar r p)}
\Bigg[ {1 \over M} \left(
g^{\alpha\beta}((1-2r)\not \! p + M)  -
{1 \over 2} (\not \! p + (1-2r)M)
[\gamma^\alpha,\gamma^\beta]
\right)
\nonumber\\
& + & 2 r \bar r {1 \over n \cdot (q - \bar r p)}
		n^\alpha  (\not \! p + M) \gamma^\beta \Bigg]{\not \! n} \; .
\label{tensor2}
\end{eqnarray}

The general procedure for extracting
the $P$-wave fragmentation functions from the above matrix elements
is the same as in the case of  the
$S$-wave \cite{bcfrags} except the intermediate steps are much more involved.
We will skip over these tedious steps and present only the final results.
{}From Eqn.(\ref{dz}), we can define
generalized fragmentation functions $D^{(1)}(z,s)$ that
depend on both the longitudinal momentum fraction $z$ and the virtuality $s$
according to
\begin{equation}
D^{(1)}(z) =
\int ds \; \theta \Biggl (s - {M^2 \over z} - {r^2 M^2 \over 1-z} \Biggr)
D^{(1)}(z,s) \; .
\label{dzs}
\end{equation}
For the $^1P_1$ state, we have
\begin{equation}
D^{(1)}_{\bar b \rightarrow \bar b c(^1P_1)} (z,s) =
{32 \alpha_s^2(2m_c) \over 81}{r \bar r^3 \over (1-\bar r z)^4}
\sum_{n=0}^3 {f_n M^{8-2n} \over (s-\bar r^2M^2)^{5-n}} \; ,
\label{dzs1p1}
\end{equation}
with
\begin{eqnarray}
f_0 &=& 64 r^2 \bar r^3 (1-\bar r z)^4 \; , \\
f_1 &=& 8 r \bar r (1- \bar r z)^3
\Biggl[  3-2r-2r^2-2\bar r (2+4r-r^2)z+\bar r^2(1-2r)z^2 \Biggr] \; , \\
f_2 &=& - (1-\bar r z)^2
\Biggl[ 2(1-2r+4r^2)-(3-42r+64r^2-16r^3)z \nonumber \\
&&  \qquad \qquad
- 2r\bar r(23-14r-4r^2)z^2+\bar r^2(1+12r)(1-2r)z^3 \Biggr] \;, \\
f_3 &=& (1-z) \Biggl[ 1-2(1-2r)z+(3-2r+2r^2)z^2
-2\bar r(2+r-2r^2)z^3+\bar r^3(2+r^2)z^4 \Biggr] \; . \nonumber \\
&& \;
\end{eqnarray}
Integrating Eqn.(\ref{dzs1p1}) over $s$ as in Eqn.(\ref{dzs}), we obtain
the fragmentation function $D^{(1)}(z)$ for the color-singlet
contribution in the $^1P_1$ state
\begin{eqnarray}
D^{(1)}_{\bar b \rightarrow \bar b c(^1P_1)} (z) & = &
{16 \alpha_s^2(2m_c) \over 243}
{r \bar r^3 z (1-z)^2 \over (1 - \bar r z)^8} \left\{
   6 - 6 (4r^2-8r+5) z \right. \nonumber \\
& + & ( 32r^4-96r^3+250r^2-210r+69) z^2 \nonumber \\
& + & 8 \bar r (4r^4+12r^3-48r^2+37r-12) z^3  \nonumber \\
& + & 2 \bar r^2 (16r^4+161r^2-114r+42) z^4
- 6 \bar r^3 (4r^3+28r^2-15r+7) z^5 \nonumber \\
& + & \left.  \bar r^4 (46r^2-14r+9) z^6 \right\} \; .
\label{dz1p1}
\end{eqnarray}
Similarly, for the $^3P_J$ states, we have
\begin{equation}
D^{(1)}_{\bar b \rightarrow \bar b c(^3P_J)} (z,s) =
{32 \alpha_s^2(2m_c) \over 243}{r \bar r^3 \over (1-\bar r z)^4}
\sum_{n=0}^3 {f^{(J)}_n M^{8-2n} \over (s-\bar r^2M^2)^{5-n}} \; ,
\label{dzs3pj}
\end{equation}
with
\begin{eqnarray}
f^{(0)}_0 &=& 64 r^2 \bar r^3 (1-\bar r z)^4 \; , \\
f^{(0)}_1 &=& 8 r \bar r (1- \bar r z)^3
\Biggl[ 1-18r+14r^2-2\bar r(1-2r+7r^2)z+\bar r^2(1+2r)z^2 \Biggr] \; , \\
f^{(0)}_2 &=& -(1-\bar r z)^2
\Biggl[ 2(1-4r)(1+6r-4r^2)-(5+14r-8r^2+80r^3-64r^4)z \nonumber \\
&&  +2\bar r(2+9r+18r^2-28r^3-16r^4)z^2
  -\bar r^2(1+6r+16r^2-32r^3)z^3 \Biggr] \;, \\
f^{(0)}_3 &=& (1-z) \Biggl[ 1-4r-(1-4r)(1-2r)z-r\bar r(3-4r)z^2 \Biggr]^2 \; ,
\end{eqnarray}
\begin{eqnarray}
f^{(1)}_0 &=& 192 r^2 \bar r^3 (1-\bar r z)^4 \; , \\
f^{(1)}_1 &=& 24 r \bar r (1- \bar r z)^3
\Biggl[ 2(1-r-r^2) - \bar r(3+10r-2r^2)z+\bar r^2 z^2 \Biggr] \; , \\
f^{(1)}_2 &=& - 6 (1-\bar r z)^2
\Biggl[ 2(1+2r) - (5-2r+6r^2)z \nonumber \\
&& \qquad \qquad + 2\bar r(2-3r-4r^2)z^2
    - \bar r^2(1-2r+2r^2)z^3 \Biggr] \;, \\
f^{(1)}_3 &=& 6(1-z) \Biggl[ 1-2(1-2r)z+(1-4r)(1-2r)z^2
+2r\bar r(1-2r)z^3+r^2 \bar r^2 z^4\Biggr] \; ,
\end{eqnarray}
and
\begin{eqnarray}
f^{(2)}_0 &=& 320 r^2 \bar r^3 (1-\bar r z)^4 \; , \\
f^{(2)}_1 &=& 8 r \bar r^2 (1- \bar r z)^3
\Biggl[ 2(4+13r)-(1+70r-26r^2)z-\bar r(7+8r)z^2 \Biggr] \; , \\
f^{(2)}_2 &=& - 4 \bar r^2 (1-\bar r z)^2
\Biggl[ 4(1+4r) -(7+12r-32r^2)z \nonumber \\
&& \qquad \qquad + 2 (1+13r-26r^2+8r^3)z^2
    + (1-30r-5r^2+4r^3)z^3 \Biggr] \;, \\
f^{(2)}_3 &=& 4 \bar r^2 (1-z) \Biggl[ 2-4(1-2r)z+(5-8r+12r^2)z^2 \nonumber\\
&& \qquad \qquad -2(1-2r)(3+2r^2)z^3+(3-12r+12r^2+2r^4)z^4\Biggr] \; .
\end{eqnarray}
Integrating Eqn.(\ref{dzs3pj}) over $s$ as in Eqn.(\ref{dzs}), we obtain
\begin{eqnarray}
D^{(1)}_{\bar b \rightarrow \bar b c(^3P_0)} (z) & = &
{16 \alpha_s^2(2m_c) \over 729}
{r \bar r^3 z (1-z)^2 \over  (1 - \bar r z)^8} \left\{
6(4r - 1)^2 + 6(4r-1)(20r^2-16r+5) z \right. \nonumber \\
& + & ( 832r^4-1456r^3+1058r^2-362r+63) z^2 \nonumber \\
& - & 8 \bar r (100r^4-184r^3+118r^2-22r+9) z^3 \nonumber \\
& + & 2 \bar r^2 (416r^4-776r^3+369r^2+42r+24) z^4 \nonumber \\
& - & 2 \bar r^3 (240r^4-516r^3+232r^2+59r+9) z^5 \nonumber \\
& + & \left.  \bar r^4 (96r^4-240r^3+134r^2+34r+3) z^6 \right\} \; ,
\label{dz3p0}
\end{eqnarray}
\begin{eqnarray}
D^{(1)}_{\bar b \rightarrow \bar b c(^3P_1)} (z) & = &
{32 \alpha_s^2(2m_c) \over 243}
{r \bar r^3 z (1-z)^2 \over (1 - \bar r z)^8} \left\{
6 + 6 (4r-5) z \right. \nonumber \\
& + & ( 16r^4+64r^2-98r+63) z^2
   + 8\bar r (2r^4+2r^3-13r^2+11r-9) z^3  \nonumber \\
& + & 2 \bar r^2 (8r^4-16r^3+47r^2-18r+24) z^4
   + 2 \bar r^3 (8r^3-24r^2+r-9) z^5 \nonumber \\
& + & \left.  \bar r^4 (12r^2+2r+3) z^6 \right\} \; ,
\label{dz3p1}
\end{eqnarray}
and
\begin{eqnarray}
D^{(1)}_{\bar b \rightarrow \bar b c(^3P_2)} (z) & = &
{64 \alpha_s^2(2m_c) \over 729}
{r \bar r^5 z (1-z)^2 \over  (1 - \bar r z)^8} \left\{
12  + 12 (2r-5) z \right. \nonumber \\
& + &  ( 92r^2-76r+135) z^2
   + 4 (10r^3-54r^2+31r-45) z^3 \nonumber \\
& + & 2 (46r^4-16r^3+123r^2-78r+75) z^4 \nonumber \\
& - & 4 \bar r (6r^4+9r^3+40r^2-13r+18) z^5 \nonumber \\
& + & \left.  \bar r^2 (12r^4-12r^3+55r^2-10r+15)  z^6 \right\} \; .
\label{dz3p2}
\end{eqnarray}
As in Ref.\cite{bcfrags},
the scale of the strong coupling constant is chosen to be
$2 m_c$ -- the minimal energy of the exchange gluon.
The results of these fragmentation functions for the $P$-wave states
disagrees with those given in Ref.\cite{pwave_chen}.

Integrating Eqns.(\ref{dz1p1}) and (\ref{dz3p0})-(\ref{dz3p2})
over $z$, we obtain the fragmentation probabilities
$P^{(1)} = (H_1/m) \int_0^1 dz \, D^{(1)}(z)$
of the color-singlet contribution for
$\bar b$ fragments into $\bar b c(n^{2S+1}P_J)$:
\begin{eqnarray}
P^{(1)}_{\bar b \rightarrow \bar b c(n^1P_1)}
  & = & {16 \alpha_s^2(2m_c) \over 243} {H_{1(\bar b c)}(n) \over m}
\Bigg[
3 {r  \log r \over \bar r^3} (10r^4+50r^3+91r^2+20r+7) \nonumber \\
&& \; \; \; \; - \; {1 \over 35 \bar r^2}
	(24r^6-256r^5-2083r^4-9538r^3-5758r^2-907r-172)
\Bigg] \; ,
\label{prob1p1}
\nonumber \\ && \;
\end{eqnarray}
\begin{eqnarray}
P^{(1)}_{\bar b \rightarrow \bar bc(n^3P_0)}
& = & {16 \alpha_s^2(2m_c) \over 729} {H_{1(\bar b c)}(n) \over m} \nonumber \\
&& \; \times \Bigg[ 3 {r  \log r \over \bar r^3}
	(32r^6+48r^5-398r^4+194r^3+337r^2+32r+1) \nonumber \\
&& \;  + \; {1 \over 35 \bar r^2}
	(9944r^6-17384r^5-34289r^4+56116r^3+11036r^2+347r+60)
\Bigg] \; ,
\label{prob3p0}
\nonumber \\ && \;
\end{eqnarray}
\begin{eqnarray}
P^{(1)}_{\bar b \rightarrow \bar bc(n^3P_1)}
  & = & {32 \alpha_s^2(2m_c) \over 243} {H_{1(\bar b c)}(n) \over m}
\Bigg[ 3 {r  \log r \over \bar r^3}
	(20r^4+46r^3+31r^2+8r+1) \nonumber \\
&& \;\;\;\; - \;
{1 \over 35 \bar r^2} (12r^6-268r^5-4461r^4-4006r^3-2326r^2-23r-58)
\Bigg] \; ,
\label{prob3p1}
\nonumber \\ && \;
\end{eqnarray}
and
\begin{eqnarray}
P^{(1)}_{\bar b \rightarrow \bar bc(n^3P_2)}
  & = & {64\alpha_s^2(2m_c) \over 729} {H_{1(\bar b c)}(n) \over m}
\Bigg[
3 {r  \log r \over \bar r^3}
	(4r^6+36r^5+5r^4-20r^3+140r^2+4r+11) \nonumber \\
&& \;\;\;\; + \; {1 \over 35 \bar r^2}
	(1468r^6+5570r^5-7345r^4+12080r^3+6130r^2+712r+285)
\Bigg] \; .
\label{prob3p2}
\nonumber \\ && \;
\end{eqnarray}

For a given principal quantum number $n$, the $^1P_1$ state
and the $^3P_1$ state, constructed in the $LS$ coupling scheme, are
mixed in general to give rise to the physical states $1^+$ and $1^{+'}$
defined as
\begin{eqnarray}
\label{mixing1}
|1^{+'}\rangle  & = & \cos \theta |^1P_1 \rangle
		\; - \; \sin \theta |^3P_1 \rangle \; , \\
|1^+ \rangle & = &  \sin \theta |^1P_1 \rangle
		\; + \; \cos \theta |^3P_1 \rangle
\; .
\label{mixing2}
\end{eqnarray}
Thus, in general, we have
\begin{eqnarray}
D^{(1)}_{\bar b \rightarrow \bar b c(1^{+'})} (z) & = &
\cos^2\theta D^{(1)}_{\bar b \rightarrow \bar b c(^1P_1)} (z) +
\sin^2\theta  D^{(1)}_{\bar b \rightarrow  \bar b c(^3P_1)} (z) -
\sin\theta \cos\theta D^{(1)}_{\rm mix} (z) \; , \\
D^{(1)}_{\bar b \rightarrow \bar b c(1^+)} (z) & = &
\sin^2\theta D^{(1)}_{\bar b \rightarrow \bar bc(^1P_1)} (z) +
\cos^2\theta  D^{(1)}_{\bar b \rightarrow  \bar b c(^3P_1)} (z) +
\sin\theta \cos\theta D^{(1)}_{\rm mix} (z) \; .
\label{dzmix}
\end{eqnarray}
Similar equations hold for the corresponding
generalized fragmentation functions
$D^{(1)}(z,s)$ as well.
With the matrix elements of ${\cal M}(n^1P_1)$ and
${\cal M}(n^3P_1)$ given by Eqns.(\ref{onepone}) -- (\ref{tensor1})
and  (\ref{matrix3pj}) -- (\ref{tensor2}) respectively, we
can calculate their interferences and obtain the mixing term:
\begin{equation}
D^{(1)}_{\rm mix} (z,s) =
{64 \sqrt 2 \alpha_s^2(2m_c) \over 81}{r \bar r^3 \over (1-\bar r z)^3}
\sum_{n=0}^2 {f^{\rm mix}_n M^{6-2n} \over (s-\bar r^2M^2)^{4-n}} \; ,
\label{dzsmix}
\end{equation}
with
\begin{eqnarray}
f^{\rm mix}_0 &=& - 8 r \bar r (1- \bar r z)^3
\Biggl[  2-3r-\bar r(1+r)z \Biggr] \; , \\
f^{\rm mix}_1 &=& (1-\bar r z)
\Biggl[ 2(1+2r^2)-(5-24r+36r^2-8r^3)z \nonumber \\
&&  \qquad \qquad
+2 \bar r (2-15r+10r^2-2r^3)z^2-\bar r^2(1-8r-4r^2)z^3 \Biggr] \;, \\
f^{\rm mix}_2 &=& -(1-z) \Biggl[ 1-(1-3r)z+r(2+r)z^2 + \bar r r^2 z^3
\Biggr] \; .
\end{eqnarray}
Integrating Eqn.(\ref{dzsmix}) over $s$ as in  Eqn.(\ref{dzs}), we obtain
\begin{eqnarray}
D^{(1)}_{\rm mix} (z) & = & - {32 \sqrt 2 \alpha_s^2(2m_c) \over 243}
		{r \bar r^3 z (1-z)^2 \over  (1 - \bar r z)^6}
\left\{   6 - 6(2r^2-4r+3) z \right. \nonumber \\
& + & (24r^3+52r^2-52r+21) z^2
  + 2 \bar r (14r^3-6r^2+15r-6) z^3 \nonumber \\
& - & \left. \bar r^2 (2r^2+8r-3) z^4 \right\} \; .
\label{dzmix1}
\end{eqnarray}
The fragmentation probability of the mixing term is given by
\begin{eqnarray}
P^{(1)}_{\rm mix} & = & {H_{1(\bar b c)}(n) \over m}
\int_0^1 dz D^{(1)}_{\rm mix} (z) \nonumber \\
  & = & - {32 \sqrt 2 \alpha_s^2(2m_c) \over 243}
{H_{1(\bar b c)}(n) \over m}
\Bigg[ 3 {r  \log r \over \bar r^3}
	(18r^4+28r^3-11r^2-2r+1)  \nonumber \\
&& \;\;\;\; + \; {1 \over 5 \bar r^2}
	(24r^5+587r^4-18r^3-88r^2-3r+8)
\Bigg] \; .
\label{probmix}
\end{eqnarray}

We next turn to the calculation of the color-octet coefficient
$D^{(8)}(z)$ to leading order in $\alpha_s$. Notice that to leading
order in $\alpha_s$, all the color-singlet contributions obtained above are
free from infrared divergences. All the long-distance effects can be factored
and absorbed into the derivative of the wavefunction, or equivalently
$H_1$ to leading order in $v^2$. The color-singlet
coefficients $D^{(1)}(z)$ do not depend on the factorization
scale $\Lambda$ to leading order in $\alpha_s$. Beyond leading order,
infrared singularities appear in the color-singlet contribution
arise from the radiation of a soft gluon in the process
$\bar b^* \to (\bar b c)\bar c g$. Imagine we are actually doing such
a next-to-leading order calculation of $D^{(1)}(z)$ with
an energy cutoff $\Lambda$ $(mv \ll \Lambda \ll m)$ imposed on
the radiated soft gluon. This cutoff will allow us
to isolate the infrared logarithmic divergent
terms in this higher order calculation and absorb them
into the nonperturbative parameter $H'_8$ associated
with the {\it color-octet mechanism}.
However,
there is a short cut to achieve the same goal by following the same
method as in Ref.\cite{gfragp}.
One can calculate the fragmentation functions for the processes
$\bar b^* \to \bar b c(^1S_0 , \underline 8)$ and
$\bar b^* \to \bar b c(^3S_1 , \underline 8)$, where the
$(\bar b c)$ pair is in the appropriate color-octet $S$-wave state.
The calculations of these fragmentation functions
are identical to the color-singlet $S$-wave case \cite{bcfrags}, except one
has to replace the $S$-wave wavefunction $R_S(0)$ by a
fictitious ``color-octet $S$-wave wavefunction at the origin"
$R_8(0)$, and the color factor $C_F^2$
by $C_F/6$ where $C_F=4/3$ for color $SU(3)$.
$R_8(0)$ is related to the nonperturbative parameter
$H'_{8(\bar b c)} (\Lambda)$ by
\begin{equation}
H'_{8(\bar b c)} (\Lambda) = {2 \over 3\pi}{|R_8(0)|^2 \over (2m)^2} \; .
\end{equation}
Using this trick, we can easily extract the coefficients of the
color-octet contributions $D^{(8)}(z)$. Analogous to Eqn.(\ref{dzs}) of the
color-singlet case, we define $D^{(8)}(z,s)$ as the following:
\begin{equation}
D^{(8)}(z) = \int ds \; \theta \Biggl( s - {M^2 \over z} -
{r^2M^2 \over 1-z} \Biggr)  D^{(8)}(z,s) \; ,
\label{dzs8}
\end{equation}
where
\begin{equation}
D^{(8)}(z,s) = {\alpha_s^2(2m_c) \over 81}{r \bar r^3 \over (1-\bar r z)^2}
\sum_{n=0}^2 {g_n M^{6-2n}  \over (s-\bar r^2 M^2)^{4-n}} \; .
\label{dzs8a}
\end{equation}
For the $^1S_0$ state, we have
\begin{eqnarray}
g_{0(^1S_0)} & = & - 12 r \bar r (1-\bar r z)^2 \; , \\
g_{1(^1S_0)} & = & - (1-\bar r z) \Biggl[ 2(1+2r)
-(1+12r-4r^2)z- \bar r (1+2r)z^2 \Biggr] \; , \\
g_{2(^1S_0)} & = & (1-z) \Biggl[ 1+2rz+(2+r^2)z^2 \Biggr] \; .
\end{eqnarray}
Similarly, for the $^3S_1$ state, we have
\begin{eqnarray}
g_{0(^3S_1)} & = & - 4 r \bar r (1-\bar r z)^2 \; , \\
g_{1(^3S_1)} & = & -(1-\bar r z) \Biggl[ 2(1-2r)
-(3-4r+4r^2)z + \bar r (1-2r)z^2 \Biggr] \; , \\
g_{2(^3S_1)} & = & (1-z) (1+rz)^2  \; .
\end{eqnarray}
Integrating Eqn.(\ref{dzs8a}) over $s$ as in Eqn.(\ref{dzs8}),
we obtain the color-octet contributions to the fragmentation functions
\begin{eqnarray}
D^{(8)}_{\bar b \to \bar b c(^1S_0)}(z) & = & {\alpha_s^2(2m_c) \over 162}
{r \bar r^3 z(1-z)^2 \over (1-\bar r z)^6}
\Bigg[  6 - 18(1 - 2r)z + (21 - 74r + 68r^2)z^2 \nonumber \\
&&\;\;\;\;\;\;\; \;\;\;
- \; 2\bar r(6 - 19r + 18r^2)z^3 + 3\bar r^2(1 - 2r + 2r^2)z^4 \Bigg] \; ,
\label{d81s0}
\end{eqnarray}
and
\begin{eqnarray}
D^{(8)}_{\bar b \to \bar b c(^3S_1)}(z) & = &
{\alpha_s^2(2m_c) \over 162}
{r \bar r^3 z(1-z)^2 \over (1-\bar r z)^6}
\Bigg[  2 - 2(3-2r)z + 3(3 - 2r + 4r^2)z^2 \nonumber \\
&&\;\;\;\;\;\;\; \;\;\;
- \; 2 \bar r(4 - r + 2r^2)z^3 + \bar r^2(3 - 2r + 2r^2)z^4 \Bigg] \, .
\label{d83s1}
\end{eqnarray}
Integrating Eqns.(\ref{d81s0}) and (\ref{d83s1})
over $z$, we obtain the fragmentation
probabilities $P^{(8)} = (H'_8/m) \int_0^1 dz \, D^{(8)}(z)$
for the color-octet contributions:
\begin{eqnarray}
P^{(8)}_{\bar b \to \bar b c(^1S_0)}
& = &
{\alpha_s^2(2m_c) \over 54} {H'_{8(\bar b c)}(\Lambda) \over m}
\Bigg[
{r\log r \over \bar r^3}(1+8r+r^2-6r^3+2r^4)
\nonumber \\
&&\;\;\;\;\;\;\; \;\;\;
+ \; {1 \over 15\bar r^2}(8+13r+228r^2-212r^3+53r^4)
\Bigg]
\; ,
\end{eqnarray}
and
\begin{eqnarray}
P^{(8)}_{\bar b \to \bar b c(^3S_1)}
& = & {\alpha_s^2(2m_c) \over 162} {H'_{8(\bar b c)} (\Lambda) \over m}
\Bigg[ {r \log r \over \bar r^3}(7-4r+3r^2+10r^3+2r^4)
\nonumber \\
&& \;\;\;\;\;\;\;\;\;\;
+ \; {1 \over 15 \bar r^2}(24+109r-126r^2+174r^3+89r^4) \Bigg]
\;.
\end{eqnarray}
To avoid large logarithms of $m/\Lambda$ appearing in the
higher order calculation of the
color-singlet coefficients $D^{(1)}(z)$,
one should choose $\Lambda \sim m$ in the matrix element
$H'_{8(\bar b c)}(\Lambda)$.

Combining both contributions from the {\it color-singlet} and
{\it color-octet mechanisms}, and including the mixing effects
in the $^1P_1$ and $^3P_1$ states, the total fragmentation probabilities are
given by
\begin{eqnarray}
P_{\bar b \to \bar b c(n^3P_0)} & = &
P^{(1)}_{\bar b \to \bar b c(n^3P_0)} + P^{(8)}_{\bar b \to \bar b c(^3S_1)}
\; , \\
P_{\bar b \to \bar b c(n \, 1^{+'})} & = &
\cos^2 \theta P^{(1)}_{\bar b \to \bar b c(n^1P_1)} +
\sin^2 \theta P^{(1)}_{\bar b \to \bar b c(n^3P_1)} -
\sin \theta \cos \theta P^{(1)}_{\rm mix} +
3 P^{(8)}_{\bar b \to \bar b c(^1S_0)} \; , \\
P_{\bar b \to \bar b c(n \, 1^+)} & = &
\sin^2 \theta P^{(1)}_{\bar b \to \bar b c(n^1P_1)} +
\cos^2 \theta P^{(1)}_{\bar b \to \bar b c(n^3P_1)} +
\sin \theta \cos \theta P^{(1)}_{\rm mix} +
3 P^{(8)}_{\bar b \to \bar b c(^3S_1)} \; , \\
P_{\bar b \to \bar b c(n^3P_2)} & = &
P^{(1)}_{\bar b \to \bar b c(n^3P_2)}
+ 5 P^{(8)}_{\bar b \to \bar bc(^3S_1)} \; .
\end{eqnarray}
As in the $S$-wave case \cite{bcfrags},
under the Altarelli-Parisi evolution to a higher scale,
the shapes of these $P$-wave fragmentation
functions are softened while the fragmentation probabilities remain unchanged.

Before closing this Section, we note that
one can also define fragmentation functions that depend on both the
longitudinal momentum fraction $z$ and the transverse momentum
$|\vec p_\bot|$. Introducing the dimensionless variable
$t = |\vec p_\bot |/M$ and using Eqn.(\ref{ptsz}) to trade the variable
$s$ with $t$, we can define the generalized fragmentation functions
$D(z,t)$ and $D(t)$ according to
\begin{eqnarray}
\int_0^\infty dt \; D(t) & = & \int_0^\infty dt \int_0^1 dz \; D(z,t) \; , \\
& = & \int ds \int dz \; \theta
\Biggl( s - {M^2 \over z} - {r^2 M^2 \over 1-z} \Biggr) D(z,s) \; .
\end{eqnarray}
Therefore,
\begin{equation}
D(z,t) = {2 M^2 t \over z(1-z)}D(z,s) \; , \;   {\rm with} \;
s = M^2 \Biggl[ {1+t^2 \over z} + {r^2 + t^2 \over 1-z} \Biggr] \; .
\end{equation}
The above relation holds for the color-singlet and the color-octet
contributions.
These functions
$D(t)$ and $D(z,t)$ are useful for studying the transverse motion of the
meson with respect to the jet axis of the fragmenting heavy quark.
The expressions of $D(t)$ for the color-octet $S$-wave
contributions can be obtained from Ref.\cite{bcfrags_pol}.
Analytic results of $D(t)$ for the $P$-wave states can also
be derived but will not be given here.

\section{Heavy quark symmetry}

In this Section, we temporarily leave
the heavy-heavy $(\bar b c)$ system and turn our attention to the
heavy-light $B$- or $D$-meson system.
In the limit of $m_Q / \Lambda_{QCD} \rightarrow \infty$, both
the heavy quark spin $\vec S_Q$
and the total spin $\vec J$ of a heavy hadron containing a
single heavy quark $Q$ become good quantum numbers.
This implies that in the spectroscopy of the hadron containing a
single heavy quark $Q$,
the angular momentum of the light degrees of freedom
$\vec J_l = \vec J - \vec S_Q$ is
also a good quantum number.  We refer collectively to
all the degrees of freedom in the heavy-light hadron
other than the heavy quark as the light degrees of freedom.
For the heavy-light $(Q \bar q)$ meson, $\vec J_l = \vec S_q + \vec L$
where $\vec S_q$ is the spin of the light antiquark $\bar q$
and $\vec L$ is the orbital angular momentum.
Thus the hadronic
states can be labeled simultaneously by the eigenvalues $j$ and $j_l$ of
the total spin $\vec J$
and the angular momentum of the light degrees of freedom
$\vec J_l$ respectively.
In general \cite{spectrum}, the spectrum of hadrons containing
a single heavy quark $Q$ has, for each $j_l$, a degenerate doublet with
total spins $j_+ = j_l + 1/2$ and $j_- =  j_l - 1/2$. (For the case of
$j_l=0$, the total spin must be $1/2$.) For $P$-wave heavy-light mesons,
the orbital angular momentum $L=1$ and $j_l$ can then be either 1/2 or 3/2.
Thus $(j_-,j_+)=(0,1)$ and (1,2) for $j_l$ = 1/2 and 3/2 respectively.
As a result, we expect to have two distinct doublets $(^3P_0,1^{+'})$ and
$(1^+,^3P_2)$ in the limit of $m_b/m_c \rightarrow \infty$, {\it i.e.}
$r \to 0$.   In this limit,
the mixing coefficients in Eqns.(\ref{mixing1})-(\ref{mixing2})
can be determined by the Clebsch-Gordon coefficients
in the tensor product of a spin $1/2$ state and a spin $1$ state
with the following result,
\begin{eqnarray}
|1^{+'} \rangle & = &   \sqrt{1 \over 3} |^1P_1 \rangle
+  \sqrt{2 \over 3} |^3P_1 \rangle \; , \\
|1^+ \rangle &  =  &
- \sqrt{2 \over 3} |^1P_1 \rangle + \sqrt{1 \over 3} |^3P_1 \rangle  \; ,
\label{mixhqlimit}
\end{eqnarray}
{\it i.e.},  we are transforming  the states
$^1P_1$ and $^3P_1$ in the $LS$ coupling scheme
to the states $1^{+'}$ and $1^+$
in the $jj$ coupling scheme.

In their discussions of the heavy quark fragmentation functions within the
context of Heavy Quark Effective Theory, Jaffe and Randall \cite{jaffe}
showed that a more natural variable to use is given by
\begin{equation}
y = {{1 \over z} - \bar r \over r} \; ,
\end{equation}
rather than the usual fragmentation variable $z$.
Furthermore, these authors also showed that the heavy quark
fragmentation function can be expanded as a power series in $r$,
\begin{equation}
D(y) = {1 \over r}a(y) + b(y) + {\cal O}(r) \; ,
\label{jafferandall}
\end{equation}
where $a(y), b(y)$, {\it etc.} are functions of the variable $y$.
The leading term $a(y)$ is constrained by the heavy quark spin-flavor
symmetry while all the higher order terms contain spin-flavor
symmetry breaking effects. One can recast our results for the $P$-wave
fragmentation functions derived in Section 3 in the above form. By
carefully expanding the powers of $r$ and $(1 - \bar r z)$,
the leading contributions to the fragmentation functions are given by
\begin{eqnarray}
D^{(1)}_{\bar b \rightarrow \bar bc(^1P_1)} (y) & \rightarrow &
\alpha_s^2 {16 \over 243r}
{(y-1)^2 \over y^8} (9y^4 - 4y^3 + 40y^2 + 96) \; , \\
D^{(1)}_{\bar b \rightarrow \bar b c(^3P_0)} (y) & \rightarrow &
\alpha_s^2 {16 \over 243r}
{(y-1)^2 \over y^8} (y^4 - 4y^3 + 8y^2 + 32) \; , \\
D^{(1)}_{\bar b \rightarrow \bar b c(^3P_1)} (y) & \rightarrow &
\alpha_s^2 {32 \over 243r}
{(y-1)^2 \over y^8} (3y^4 - 4y^3 + 16y^2 + 48) \; , \\
D^{(1)}_{\bar b \rightarrow \bar bc(^3P_2)} (y) & \rightarrow &
\alpha_s^2 {320 \over 243r}
{(y-1)^2 \over y^8} (y^4 + 4y^2 + 8) \; , \\
D^{(1)}_{\rm mix} (y) & \rightarrow &
- \alpha_s^2 {32 \sqrt 2 \over 243r}
{(y-1)^2 \over y^6} (3y^2 + 4y + 8) \; .
\end{eqnarray}
Therefore, in the heavy quark limit, we have
\begin{eqnarray}
D^{(1)}_{\bar b \rightarrow \bar bc(1^{+'})} (y) & \rightarrow &
{1 \over 3}  D^{(1)}_{\bar b \rightarrow  \bar b c(^1P_1)} (y) +
{2 \over 3} D^{(1)}_{\bar b \rightarrow \bar b c(^3P_1)} (y) +
{\sqrt 2 \over 3} D^{(1)}_{\rm mix} (y) \; , \nonumber\\
& \rightarrow & \alpha_s^2 {16 \over 81r}
{(y-1)^2 \over y^8} (y^4 - 4y^3 + 8y^2 + 32) \; , \\
D^{(1)}_{\bar b \rightarrow \bar bc(1^+)} (y) & \rightarrow &
{2 \over 3}  D^{(1)}_{\bar b \rightarrow  \bar b c(^1P_1)} (y) +
{1 \over 3} D^{(1)}_{\bar b \rightarrow \bar bc(^3P_1)} (y) -
{\sqrt 2 \over 3} D^{(1)}_{\rm mix} (y) \; , \nonumber\\
& \rightarrow & \alpha_s^2 {64 \over 81r}
{(y-1)^2 \over y^8} (y^4 + 4y^2 + 8) \; .
\end{eqnarray}
These imply the following spin counting ratios
\begin{eqnarray}
{D^{(1)}_{\bar b \rightarrow \bar bc(^3P_0)} (y) \over
 D^{(1)}_{\bar b \rightarrow \bar bc(1^{+'})} (y) }
&  \rightarrow & {1 \over 3} \; , \\
{D^{(1)}_{\bar b \rightarrow \bar bc(1^+)} (y) \over
 D^{(1)}_{\bar b \rightarrow \bar bc(^3P_2)} (y) }
&  \rightarrow & {3 \over 5}
\; ,
\end{eqnarray}
as expected from the heavy quark symmetry.

Similarly, as already shown in
Ref.\cite{bcfrags}, the color-octet coefficients
$D^{(8)}_{\bar b \to \bar b c(^1S_0)}(z)$
and $D^{(8)}_{\bar b \to \bar b c(^3S_1)}(z)$ given by
Eqn.(\ref{d81s0}) and (\ref{d83s1}) respectively
also satisfy the heavy
quark spin symmetry in the limit of $r \to 0$.
Thus, heavy quark spin symmetry is a powerful
tool to check our tedious results obtained in previous Section.
We note that the mixing effects and the color-octet contributions were not
considered in Ref.\cite{pwave_chen}.  Moreover, the results given there
do not have the proper heavy quark limit given by Eqn.(\ref{jafferandall}).

Since our P-wave fragmentation functions derived in previous
Section are consistent with the general QCD analysis of heavy quark
fragmentation functions  by Jaffe and Randall \cite{jaffe},
they may be useful as phenomenological fragmentation functions for
$c \to D^{**}$ and $b \to B^{**}$, with the mass ratio $r$ and the two
overall factors $\alpha_s^2 H_1/m$ and $\alpha_s^2 H'_8/m$
treated as free parameters.

\bigskip

\section{Braaten-Levin spin counting rule}

The fragmentation function $D_{i \to H}(z)$ for a parton $i$ splitting into a
hadron $H$ is related to the distribution function $f_{i/H}(x)$
of finding the parton $i$ inside the hadron $H$
by analytic continuation \cite{drelletal,lipatovetal}
\begin{equation}
f_{i/H}(x) = x D_{i \to H}({1 \over x}) \; .
\label{continuation}
\end{equation}
The results of our perturbative $S$-wave and $P$-wave fragmentation functions
obtained in Ref.\cite{bcfrags,charm} and in this paper allow us to study
the perturbative tail of the distribution functions of the heavy quark
inside the heavy $S$-wave and $P$-wave mesons as well. From our explicit
calculations, we see that $f_{i/H}(x)$ has a pole located at $x=\bar r$.
The pole is cut off by nonperturbative effects related to the formation of
bound states of the $(\bar b c)$ pair.
This pole is of order 6 and 8 for the $S$-wave and $P$-wave states
respectively. In the general case of $L$-waves
($L=0,1,2,\cdots$ or $S$-, $P$-, $D$-waves, $\cdots$),
we expect this pole is of order $6+2L$. Therefore, we can expand $f(x)$ as a
Laurent series,
\begin{equation}
f(x) = {a_{n}(r) \over (x - \bar r)^n} \; + \;
{\rm less \; singular \; terms} \; ,
\end{equation}
with $n=6+2L$ for the general $L$-waves.

An interesting result recently obtained
by Braaten and Levin \cite{eric}
states that the $r$-dependent coefficients ${a_n(r)}$ satisfy
the simple spin counting rules. To be more specific, we have
\begin{equation}
a_6(^1S_0):a_6(^3S_1) \; = \; 1:3 \; ,
\end{equation}
for the $S$-waves, and
\begin{equation}
a_8(^1P_1):a_8(^3P_0):a_8(^3P_1):a_8(^3P_2) \; = \;
3:1:3:5 \; ,
\end{equation}
for the $P$-waves.

{}From Eqns.(\ref{d81s0}) and (\ref{d83s1}) or using the results in
Ref.\cite{bcfrags}, we can calculate the two coefficients $a_6(^1S_0)$
and $a_6(^3S_1)$  for the $S$-wave states to leading order in $\alpha_s$
(up to an overall common factor of $\alpha_s^2 H_8'/m$),
\begin{eqnarray}
a_6(^1S_0) & = & \lim_{x \to \bar r}
\Biggl[ (x - \bar r)^6 x D^{(8)}_{^1S_0}({1 \over x}) \Biggr] =
{4  \over 81}(r \bar r)^5 \; , \\
a_6(^3S_1) & = & 3 \lim_{x \to \bar r}
\Biggl[ (x - \bar r)^6 x D^{(8)}_{^3S_1}({1 \over x}) \Biggr] =
{4  \over 27}(r \bar r)^5 \; .
\end{eqnarray}
Similarly, from Eqns.(\ref{dz1p1}), (\ref{dz3p0}), (\ref{dz3p1}), and
(\ref{dz3p2}), we obtain the coefficients $a_8(^1P_1)$ and $a_8(^3P_J)$
for the $P$-wave states to leading order in $\alpha_s$
(up to an overall common factor of $\alpha_s^2 H_1/m$),
\begin{eqnarray}
a_8(^1P_1) & = & \lim_{x \to \bar r}
\Biggl[ (x - \bar r)^8 x D^{(1)}_{^1P_1}({1 \over x}) \Biggr] =
{512  \over 81}(r \bar r)^7 \; , \\
a_8(^3P_0) & = & \lim_{x \to \bar r}
\Biggl[ (x - \bar r)^8 x D^{(1)}_{^3P_0}({1 \over x}) \Biggr] =
{512  \over 243}(r \bar r)^7 \; , \\
a_8(^3P_1) & = & \lim_{x \to \bar r}
\Biggl[ (x - \bar r)^8 x D^{(1)}_{^3P_1}({1 \over x}) \Biggr] =
{512  \over 81}(r \bar r)^7 \; , \\
a_8(^3P_2) & = & \lim_{x \to \bar r}
\Biggl[ (x - \bar r)^8 x D^{(1)}_{^3P_2}({1 \over x}) \Biggr] =
{2560  \over 243}(r \bar r)^7 \; .
\end{eqnarray}
Moreover, from Eqn.(\ref{dzmix1}), we have
\begin{equation}
a_8^{\rm mix}  =  \lim_{x \to \bar r}
\Biggl[ (x - \bar r)^8 x D^{(1)}_{\rm mix}({1 \over x}) \Biggr] = 0 \; .
\end{equation}
Therefore, our explicit results of the $S$-wave and $P$-wave fragmentation
functions do satisfy the Braaten-Levin spin counting rule.
Notice that
unlike the spin counting in the heavy quark limit studied in the last Section
which requires $r \to 0$, the Braaten-Levin
spin counting rule holds for arbitrary values of $r$.

\bigskip

\section{Fragmentation functions for $c \to h_c$ and
$c \to \chi_{cJ} \; (J = 0,1,2)$}

Our results derived in Sections 2 and 3 can be easily applied to the cases of
$P$-wave charmonium and bottomonium
by taking the equal mass limit of $m_c=m_b$,  {\it i.e.}, $r=\bar r=1/2$.
Simple formulas can be obtained in this limit and are given below
for convenience. To be specific, we will consider the $P$-wave charmonium case.
To leading order in $v^2$,
the fragmentation functions for a charm (or anti-charm) quark  to
fragment into various $P$-wave charmonium states consist of two terms:
\begin{eqnarray}
D_{c \to h_c}(z,\mu_0) & = & {H_{1(c \bar c)} \over m_c}
D^{(1)}_{c \to c \bar c(^1P_1)}(z,\Lambda) +
3 {H'_{8 (c \bar c)}(\Lambda) \over m_c}
D^{(8)}_{c \to c \bar c(^1S_0)}(z) \; , \\
D_{c \to \chi_{cJ}}(z,\mu_0) & = & {H_{1(c \bar c)} \over m_c}
D^{(1)}_{c \to c \bar c(^3P_J)}(z,\Lambda)
+ (2J+1) {H'_{8(c \bar c)} (\Lambda) \over m_c}
D^{(8)}_{c \to c \bar c(^3S_1)}(z) \; ,
\end{eqnarray}
where $\Lambda$ is a factorization scale within the range
$m_c v \ll \Lambda \ll m_c$.
$H_{1(c \bar c)}$ is related to the derivative of the radial wavefunction
of the $P$-wave charmonium, $R^{\prime}_P(0)$,
according to \cite{bbl_long,bbl_short}
\begin{equation}
H_{1(c \bar c)} \approx {9 \over 2 \pi} {|R^{\prime}_P(0)|^2 \over m_c^4} \; .
\end{equation}
$H'_{8(c \bar c)}$ satisfies the
renormalization group equation \cite{bbl_long,bbly}
\begin{equation} {
\Lambda {d \ \over d \Lambda} H'_{8(c \bar c)}(\Lambda)
\; = \; {16 \over 27\pi}
	\alpha_s(\Lambda)H_{1(c \bar c)} \; ,
} \label{rgeqccbar} \end{equation}
with the solution
\begin{equation} {
H'_{8(c \bar c)}(\Lambda) \;  = \; H'_{8(c \bar c)} (\Lambda_0)
\; + \; {16 \over 27 \beta_0}
	\log \left( {\alpha_s(\Lambda_0) \over \alpha_s(\Lambda)} \right)
\; H_{1(c \bar c)} \; ,
} \label{rgsolccbar} \end{equation}
where $\beta_0 = 25/6$ is the first coefficient in the beta function for QCD
with 4 light flavors.
If the factorization scale $\Lambda$ is chosen to be much less than
the charm quark mass $m_c$,
the above equation can be used to sum up large logarithms of $m_c/\Lambda$.

To leading order in $\alpha_s$, we have
\begin{eqnarray}
D^{(1)}_{c \to c\bar c(^1P_1)}(z) & = &
{16 \alpha_s^2(2m_c) \over 81}
{z(1-z)^2 \over (2-z)^8} \nonumber \\
&& \times \Bigg[ 64 - 128z + 176z^2 - 160z^3 + 140z^4 - 56z^5 + 9z^6 \Bigg]
\; , \\
D^{(1)}_{c \to c\bar c(^3P_0)}(z) & = &
{16 \alpha_s^2(2m_c) \over 729}
{z(1-z)^2 \over (2-z)^8}\nonumber \\
&& \times
\Bigg[ 192 + 384z + 528z^2 - 1376z^3 + 1060z^4 - 376z^5 + 59z^6 \Bigg]
\; ,  \\
D^{(1)}_{c \to c\bar c(^3P_1)}(z) & = &
{64 \alpha_s^2(2m_c) \over 243}
{z(1-z)^2 \over (2-z)^8} \nonumber \\
&& \times \Bigg[ 96 - 288z + 496z^2 - 408z^3 + 202z^4 - 54z^5 + 7z^6 \Bigg]
\; , \\
D^{(1)}_{c \to c\bar c(^3P_2)}(z) & = &
{128 \alpha_s^2(2m_c) \over 729}
{z(1-z)^2 \over (2-z)^8} \nonumber \\
&& \times \Bigg[ 48 - 192z + 480z^2 - 668z^3 + 541z^4 - 184z^5 + 23z^6 \Bigg]
\; .
\label{pwavecharmdz}
\end{eqnarray}
The corresponding fragmentation probabilities
$P^{(1)} = (H_1/m_c) \int_0^1 dz \, D^{(1)} (z)$
are simply given by
\begin{eqnarray}
P^{(1)}_{c \to h_c} & = &
{8 \alpha_s^2(2m_c) \over 81} {H_{1(c \bar c)} \over m_c}
\Bigg[ {18107 \over 35} - 746 \log 2 \Bigg] \; , \\
P^{(1)}_{c \to \chi_{c0}} & = &
{8 \alpha_s^2(2m_c) \over 729}{H_{1(c \bar c)} \over m_c}
\Bigg[ {119617 \over 35} - 4926 \log 2 \Bigg] \; , \\
P^{(1)}_{c \to \chi_{c1}} & = &
{64 \alpha_s^2(2m_c) \over 243}{H_{1(c \bar c )} \over m_c}
\Bigg[ {1151 \over 7} - 237 \log 2 \Bigg] \; , \\
P^{(1)}_{c \to \chi_{c2}} & = &
{32 \alpha_s^2(2m_c) \over 729}{H_{1(c \bar c)} \over m_c}
\Bigg[ {54743 \over 35} - 2256 \log 2 \Bigg] \; .
\label{pwavecharmprob}
\end{eqnarray}

The color-octet coefficients can be obtained
by making appropriate changes in the radial wavefunction and color factor
to the results of the $S$-wave fragmentation functions
$D_{c \to \eta_c}(z)$ and $D_{c \to J/\psi}(z)$ given in Ref.\cite{charm},
or simply by setting $r=\bar r=1/2$ in Eqns.(\ref{d81s0}) and (\ref{d83s1}).
The results are
\begin{eqnarray}
D^{(8)}_{c \to c\bar c(^1S_0)}(z) & = & {\alpha_s^2(2m_c) \over 162}
{z(1-z)^2 \over (2-z)^6}
\Bigg[ 48 + 8z^2 - 8z^3 +3 z^4 \Bigg] \; , \\
D^{(8)}_{c \to c \bar c(^3S_1)}(z) & = & {\alpha_s^2(2m_c) \over 162}
{z(1-z)^2 \over (2-z)^6}
\Bigg[ 16 - 32z + 72z^2 -32z^3 + 5z^4 \Bigg] \; .
\end{eqnarray}
The fragmentation probabilities
$P^{(8)} = (H'_8/m_c) \int_0^1 dz \, D^{(8)} (z)$
of the color-octet piece are given by
\begin{eqnarray}
P^{(8)}_{c \to c \bar c(^1S_0)} & = &
{\alpha_s^2(2m_c) \over 54}{H'_{8(c \bar c)} (\Lambda) \over m_c}
\Bigg[ {773 \over 30} - 37 \log 2 \Bigg] \;\; , \\
P^{(8)}_{c \to c \bar c(^3S_1)} & = &
{\alpha_s^2(2m_c) \over 162}{H'_{8(c \bar c)} (\Lambda) \over m_c}
\Bigg[ {1189  \over 30} - 57 \log 2 \Bigg] \; \; .
\end{eqnarray}
To avoid large logarithms of $m_c/\Lambda$ appearing in the color-singlet
contributions, one should set $\Lambda \sim m_c$ in the matrix element
$H'_{8(c \bar c)}(\Lambda)$.

Adding both color-singlet and color-octet contributions, the total
fragmentation probabilities for $c$ or $\bar c$ to split into the 4 $P$-wave
charmonium state are given by
\begin{eqnarray}
P_{c \to h_c} & = &
P^{(1)}_{c \to c \bar c(^1P_1)} + 3 P^{(8)}_{c \to c \bar c(^1S_0)} \; , \\
\label{factor5}
P_{c \to \chi_{cJ}} & = &
P^{(1)}_{c \to c \bar c(^3P_J)} + (2J+1)P^{(8)}_{c \to c \bar c(^3S_1)} \; .
\label{factor6}
\end{eqnarray}
We note that in the equal mass case, mixings between the $^1P_1$ state and
the $^3P_1$ state are not allowed by charge conjugation.
Once produced by fragmentation, the $P$-wave $\chi_{cJ}$ states
radiatively decay into $J/\psi$ and therefore contribute to the
inclusive $J/\psi$ production cross section.

Extensions of the formulas given in this Section to
the $P$-wave states of bottomonium system are straightforward.

\section{Discussions}

According to the results from potential model
calculations \cite{eichtenquigg}, the radial wavefunctions
at the origin for the first two sets of $P$-wave $(\bar b c)$ bound states
are given by
$|R'_{2P}(0)|^2 = 0.201 \; {\rm GeV}^5$ and
$|R'_{3P}(0)|^2 = 0.264 \; {\rm GeV}^5$.
The corresponding
$H_{1 (\bar b c)}(2)$ and $H_{1 (\bar b c)}(3)$ are about 10 MeV and 14 MeV
respectively, where we have used $m_c$ = 1.5 GeV and $m_b$ = 4.9 GeV.
We will choose the factorization scale $\Lambda = m$.
In the limit of $m \gg \Lambda_0$, the contribution to $H'_{8(\bar b c)}$
from the perturbative evolution dominates, and one can estimate
$H'_{8(\bar b c)}(m)$ by setting $\alpha_s(\Lambda_0) \sim 1$ and
neglecting the constant $H'_{8(\bar b c)}(\Lambda_0)$ in
Eqn.(\ref{rgsol}). Taking $\alpha_s(m) = 0.38$, we estimate
$H'_{8(\bar b c)}(m) \approx 1.3 \; {\rm MeV}$ and $1.8 \; {\rm MeV}$
for the $2P$ and $3P$ states respectively.  Numerically, the
color-octet contributions are small compared with the color-singlet
contributions.
The mixing angle for the
$^1P_1$ and $^3P_1$ states are also obtained in \cite{eichtenquigg} with
the results
$(\cos\theta,\sin\theta)$ = (0.999,0.030) and
(0.957,0.290) for the $2P$ and $3P$ states respectively.  The mixing angle
is surprisingly small which implies the states constructed in the
$LS$ coupling scheme are actually very close to the physical eigenstates.
The total fragmentation probabilities for
$\bar b$ to split into the four $2P$ states are estimated to be
\begin{eqnarray}
\label{twop0}
P_{\bar b \to \bar b c(2^3P_0)}  \approx 2.3 \times 10^{-5} \; , \\
\label{twop1pi}
P_{\bar b \to \bar b c(2\,1^{+'})} \approx 4.4 \times 10^{-5} \; , \\
\label{twop1}
P_{\bar b \to \bar b c(2\,1^+)}    \approx 4.8 \times 10^{-5} \; , \\
P_{\bar b \to \bar b c(2^3P_2)}  \approx 5.6 \times 10^{-5} \; .
\label{twop2}
\end{eqnarray}
Similarly, for the $3P$ states, we obtain
\begin{eqnarray}
\label{threep0}
P_{\bar b \to \bar b c(3^3P_0)}  \approx 3.0 \times 10^{-5} \; , \\
\label{threep1pi}
P_{\bar b \to \bar b c(3\,1^{+'})} \approx 8.1 \times 10^{-5} \; , \\
\label{threep1}
P_{\bar b \to \bar b c(3\,1^+)}    \approx 4.0 \times 10^{-5} \; , \\
P_{\bar b \to \bar b c(3^3P_2)}  \approx 7.4 \times 10^{-5} \; .
\label{threep2}
\end{eqnarray}

Since all the $2P$ states will decay $100 \%$ to the $1S$ pseudoscalar
ground state $B_c$,  one should add up all the probabilities from
Eqns.(\ref{twop0})-(\ref{twop2}) to give the total fragmentation probability
for the $B_c$ production rate from the cascades of the $2P$ states, which
is about $1.7 \times 10^{-4}$. This is comparable to the
probability $3.8 \times 10^{-4}$ of the direct fragmentation of
$\bar b \to B_c$, and is about 10 \% of the  lower bound
$1.5 \times 10^{-3}$ for the probability
of $\bar b \to B_c$ including the $1S$ and $2S$ states
obtained in Ref.\cite{bcfrags}.
Similarly, the total probability for
the production of $B_c$ from the cascades of the $3P$ states
is about $2.3 \times 10^{-4}$.
Therefore
the $2P$ and $3P$ states together account for
a significant fraction of the inclusive
production rate of the $B_c$.
While the $2P$ states are expected to lie below the $BD$ flavor threshold,
the $3P$ states may or may not lie below this threshold. If
any one of the four
$3P$ states lies above this flavor threshold, it will quickly dissociate
into a pair of $B$ and $D$ mesons, and will not contribute to the
inclusive $B_c$ production rate.
The $p_T$ distributions and the total inclusive production
cross sections for these $P$-wave states with realistic rapidity cut
at the $p \bar p$ colliders are now under
investigation \cite{bcp_tevatron}.
The four $3D$ states are also expected
to lie below the $BD$ threshold \cite{eichtenquigg}.
The corresponding $D$-wave fragmentation functions
and probabilities have not yet been calculated.

For the $P$-wave charmonium case, $H_{1(c \bar c)}$ has been
phenomenologically determined to be
$\approx 15$ MeV from the light hadronic decay rates of the $\chi_{c1}$ and
$\chi_{c2}$ \cite{bbl_short}. We set the factorization scale $\Lambda = m_c$.
Following Ref.\cite{gfragp},
we will take $H'_{8(c \bar c)}(m_c) \approx 3$ MeV. Numerically,
the color-octet contributions are minuscule.
The total fragmentation probabilities for $c \to h_c$ and $c \to \chi_{cJ}$
are estimated to be
\begin{eqnarray}
\label{phc}
P_{c \to h_c}       \approx 1.8 \times 10^{-5} \; , \\
\label{pchi0}
P_{c \to \chi_{c0}} \approx 2.4 \times 10^{-5} \; , \\
\label{pchi1}
P_{c \to \chi_{c1}} \approx 2.8 \times 10^{-5} \; , \\
\label{pchi2}
P_{c \to \chi_{c2}} \approx 1.1 \times 10^{-5} \; .
\end{eqnarray}
The production of $\chi_{cJ}$ states contribute to the inclusive
production rate of $J/\psi$ through the radiative decay
$\chi_{cJ} \to J/\psi + \gamma$. Multiplying the fragmentation probabilities
given above by the appropriate branching ratios of
$0.7\%,27\%,$ and $14\%$, we find that the probability of a $J/\psi$ inside a
primary charm quark jet coming from the cascades of the $\chi_{cJ}$ states
is approximately $0.9 \times 10^{-5}$.
This is about an order of magnitude smaller than the probability
$1.2 \times 10^{-4}$
for the direct fragmentation of $c \to J/\psi$ obtained in Ref.\cite{charm}
but still larger than
the probability $3 \times 10^{-6}$  for the direct fragmentation
of $g \to J/\psi$ obtained in Ref.\cite{gfrags}.
The probability of finding a $J/\psi$ inside a
primary gluon jet coming from the cascades of the $\chi_{cJ}$ states
has also been estimated to be about $8 \times 10^{-5}$ \cite{gfragp}.
The inclusive $J/\psi$ production rate at large transverse momentum region
in $p \bar p$ collisions is
expected to be dominated by the gluon fragmentation into
$\chi_{cJ}$ followed by their radiative decays into $J/\psi$
\cite{gfragp}.
Fragmentation contributions to the inclusive
production cross sections of $J/\psi$, $\psi'$, and $\chi_{cJ}$
at  $p \bar p$ colliders
are now under investigation \cite{flemingetal}.

Fragmentation is the dominant
mechanism for production of $(\bar b c)$ mesons
at large transverse momentum in high energy colliders.
In this paper, we have calculated the
fragmentation functions for production
of the $P$-wave $(\bar b c)$ mesons to leading order in $\alpha_s$.
These $P$-wave fragmentation functions can be used to calculate the
inclusive $P$-wave $(\bar b c)$ mesons production cross sections
at large transverse momentum. The production of the $P$-wave states account
for about 20~\% of the inclusive production rate of the ground state
$B_c$.
The $D$-wave contributions to the inclusive production rate of $B_c$
should be smaller than those of the $P$-waves.
Thus these $P$-wave fragmentation functions, together with the $S$-wave
fragmentation functions obtained in Ref.\cite{bcfrags}, should allow an
accurate calculation of the production rate of the $B_c$ meson at large
transverse momentum in high energy colliders \cite{bcp_tevatron}.

\vfill\eject

I am grateful to Eric Braaten for useful discussions and
reading the manuscript.  I would also like to thank
Dr. Heath Pois for checking some of the results presented in this work.
This work was supported in part by the U.S. Department of Energy,
Division of High Energy Physics under Grants DE-FG03-91ER40674
and by Texas National Research Laboratory Grant RGFY93-330.

\vfill\eject

\bigskip
\noindent{\Large\bf Figure Caption}
\begin{enumerate}
\item Feynman diagram for $\bar b^* \rightarrow \bar b c \bar c$
which contributes to the fragmentation of $\bar b$ into $P$-wave $(\bar b c)$
bound states.  The outgoing momenta are $(1-r)p+k$, $rp-k$, and $p'$ for the
$\bar b$, $c$, and $\bar c$, respectively. $k$ is the relative momentum of the
$\bar b$ and $c$.
\end{enumerate}
\vfill\eject

\end{document}